# Broken symmetry dielectric resonators for high quality-factor Fano metasurfaces

Salvatore Campione,[1,2,3,*] Sheng Liu,[1,2,3] Lorena I. Basilio,[1] Larry K. Warne,[1] William L. Langston,[1] Ting S. Luk,[1,2] Joel R. Wendt,[1] John L. Reno,[1,2] Gordon A. Keeler,[1] Igal Brener,[1,2] Michael B. Sinclair,[1,*]

[1]Sandia National Laboratories, P.O. Box 5800, Albuquerque, NM 87185, USA
[2]Center for Integrated Nanotechnologies, Sandia National Laboratories, P.O. Box 5800, Albuquerque, NM 87185, USA
[3]S.C. and S.L. contributed equally
*Corresponding authors: sncampi@sandia.gov or mbsincl@sandia.gov

We present a new approach to dielectric metasurface design that relies on a single resonator per unit cell and produces robust, high quality-factor Fano resonances. Our approach utilizes symmetry breaking of highly symmetric resonator geometries, such as cubes, to induce couplings between the otherwise orthogonal resonator modes. In particular, we design perturbations that couple "bright" dipole modes to "dark" dipole modes whose radiative decay is suppressed by local field effects in the array. Our approach is widely scalable from the near- infrared to radio frequencies. We first unravel the Fano resonance behavior through numerical simulations of a germanium resonator-based metasurface that achieves a quality-factor of ~1300 at ~10.8 µm. Then, we present two experimental demonstrations operating in the near-infrared (~1 µm): a silicon-based implementation that achieves a quality-factor of ~350; and a gallium arsenide-based structure that achieves a quality-factor of ~600 — the highest near-infrared quality-factor experimentally demonstrated to date with this kind of metasurfaces. Importantly, large electromagnetic field enhancements appear within the resonators at the Fano resonant frequencies. We envision that combining high quality-factor, high field enhancement resonances with nonlinear and active/gain materials such as gallium arsenide will lead to new classes of active optical devices.

Metasurfaces are currently the subject of intensive research worldwide since they can be tailored to produce a wide range of optical behaviors. However, metasurfaces generally exhibit broad spectral resonances, and it is difficult to obtain narrow (i.e. high quality-factor, Q) spectral features. Attaining such high-Q features from metasurfaces would greatly expand their application space, particularly in the areas of sensing, spectral filtering, and optical modulation. Early metasurfaces were fabricated from metals and exhibited particularly broad resonances at infrared and optical frequencies as a result of Ohmic losses. Dielectric resonator-based metasurfaces were introduced to overcome these losses and have enabled, among others, wave-front manipulation and cloaking devices, perfect reflectors, and ultrathin lenses [1-10] but, although absorptive losses were reduced, the metasurface resonances remained broad due to strong coupling with the external field (i.e. large radiation losses).

Recently, new strategies based on "electromagnetically induced transparency" or "Fano resonances" have been developed that show great promise for achieving high-Q resonances [11-15]. In this approach, the resonator system is designed to support both "bright" and "dark" resonances. The incident optical field readily couples to the bright resonance, but cannot couple directly to the dark resonance. Through proper design, a weak coupling between the two resonances can be introduced, allowing energy from the incident wave to be indirectly coupled to the dark resonance. The metasurface transmission and reflection spectra resulting from such an approach feature Fano resonances that can be much narrower than the traditional metasurface resonances. This approach has been demonstrated for metal-based metasurfaces at THz frequencies where Q-factors approaching 100 have been observed [11, 12]. Even more dramatic results have been achieved by applying this strategy to dielectric resonator-based metasurfaces and Q-factors approaching 500 have been demonstrated [14].

A common feature of the dielectric resonator-based Fano designs demonstrated thus far is the reliance on multiple, distinct, near-field coupled dielectric structures within the unit cell [14-16]. However, reliable and repeatable control of near-field coupling requires exacting fabrication tolerances and it is of interest to inquire if simpler high-Q Fano designs can be developed. In this paper, we present a new monolithic resonator design that achieves ultra-high Q-factors while using only one dielectric structure per unit cell. Our approach relies on breaking the symmetry of otherwise highly symmetric resonators to induce intra-resonator mixing of bright and dark modes (rather than inter-resonator couplings as in Refs. [14-16] ), and is scalable from the near- infrared to radio frequencies. A related approach has recently been demonstrated that uses non-resonant scatterers to directly excite dark modes of very thin dielectric disks [17]. We begin by presenting a series of numerical simulations for a germanium (Ge)-based Fano design to unravel the origins of the high-Q Fano resonances. Next, we exploit the wavelength scalability of our approach and present two experimental demonstrations of metasurfaces with extremely sharp transmission resonances in the near- infrared (~1 µm). The first utilizes resonators fabricated on a silicon-on-insulator (SOI) substrate and achieves a Q-factor of about 350. The second experimental demonstration utilizes gallium arsenide (GaAs) resonators and achieves a Q-factor of about 600. To our knowledge, this is the highest Q-factor reported to date for this kind of metasurfaces operating in this spectral range. The achievement of high quality factors in the infrared spectral range (thanks to the scalability of our design) would enable biosensing devices and tunable narrowband passband filters using liquid crystals, for example. Importantly, extending this approach to an active and nonlinear material such as GaAs leads to new device opportunities in areas such as lasers, detectors, modulators, and sensors.

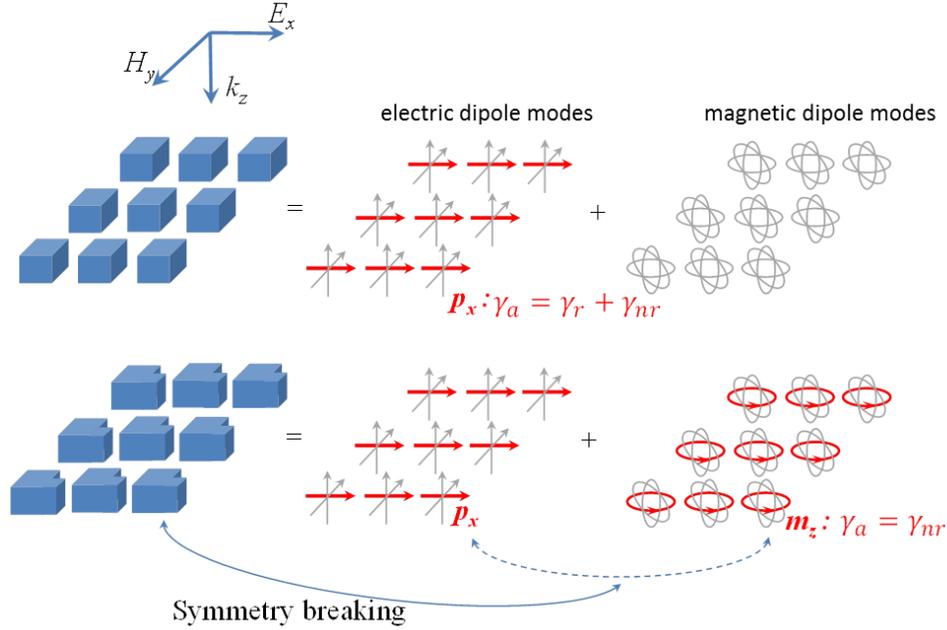

Fig. 1. A schematic depiction of the operating principles of the Fano metasurfaces. The top row shows that the electric field of the incident radiation excites only the $p_x$ electric dipole mode of the cube resonators (red arrows). The decay of the electric dipole mode is governed by both radiative ($r$) and non-radiative ($nr$) processes. In the bottom row, symmetry breaking has been used to allow coupling of the $p_x$ dipole to the longitudinal $m_z$ magnetic dipole (red loops) which only decays due to non-radiative losses. A similar process (not shown) involving the bright in-plane magnetic dipole ($m_y$) and dark longitudinal electric dipole ($p_z$) leads to a second Fano resonance at higher energy.

The principles underlying our high-Q Fano metasurfaces are shown schematically in Fig. 1. We start with a simple cubic resonator similar to the dielectric resonators demonstrated in Ref. [1]. For an isolated resonator, such a high-symmetry geometry leads to orthogonal, but degenerate, sets of electric and magnetic dipole modes oriented along the $x$–, $y$–, and $z$–directions (along with other higher order multipoles). When arranged in an array with subwavelength periodicity, only the transverse (i.e. in-plane) dipole modes can couple to a normally incident electromagnetic wave (Fig. 1, top row) and this results in the usual (broad) electric and magnetic transmission/reflection resonances [1]. However, it is possible to "perturb" the geometry to change the spectral positions of the modes [18-20], or even to induce mode mixing between the transverse and longitudinal dipole modes. The bottom row of Fig. 1 shows a symmetry breaking induced coupling between the $p_x$ electric dipole mode and the longitudinal $m_z$ magnetic dipole mode. While the $p_x$ dipole is subject to both radiative and non-radiative decay processes, the $m_z$ mode is subject to only non-radiative decay and high Q-values can be achieved using low loss dielectric materials. The interference between these two modes leads to the observed high-Q Fano resonances. A similar process (not shown) involving the bright in-plane magnetic dipole ($m_y$) and dark longitudinal electric dipole ($p_z$) leads to a second Fano resonance at higher energy.

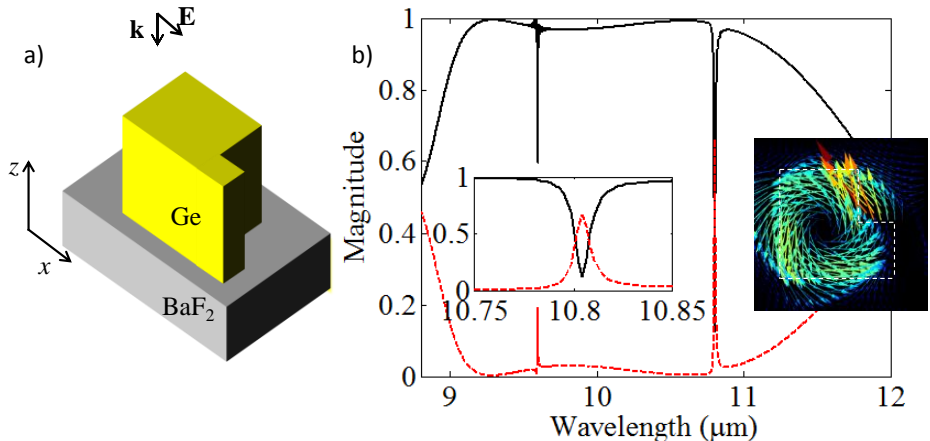

Fig. 2. (a) A schematic of the unit cell of the Fano metasurface design. The period of the metasurface is 4.2 µm. (b) The reflectivity (solid black) and the transmissivity (dashed red) spectra of the array in part (a) obtained using full-wave simulations. The inset on the left shows an expanded view of the resonance at ~10.8 µm. The Q-factor for this resonance is ~1300. The inset on the right shows a vector plot of the electric field in the $x-y$ resonator mid-plane featuring a circulating pattern.

Figure 2(a) shows a schematic of such a broken symmetry resonator design that utilizes germanium as the resonator material. Starting with a cubic geometry of nominal side length 2.53 µm, a small notch is cut from one corner of the cube (the details of the resonator design are presented in the Supplementary material). The resonators are arrayed on a barium fluoride (BaF2) substrate with an array period of 4.2 µm. Such a low index substrate is necessary to retain the original Mie modes of the dielectric resonators [1]. Figure 2(b) shows the reflectivity and the transmissivity spectra of the array under $x$–polarized incidence obtained from a Finite Difference Time Domain (FDTD) simulation (FDTD Solutions, Lumerical, Inc.). Several extremely narrow Fano resonances are observed — the transmissivity spectrum exhibits pass-band resonances. Note that the Ge properties used in the simulations included the appropriate material absorptive loss values. The quality-factor of the Fano reflection resonance (as defined by $\lambda_0/\Delta\lambda$ where $\lambda_0$ is the resonant frequency and $\Delta\lambda$ is the full width at half minimum (FWHM) of the resonance) at ~10.8 μm exceeds 1300. Furthermore, at the Fano resonant frequency, the electric and magnetic fields within the resonator are enhanced by several orders of magnitude relative to the incident field.

The inset of Fig. 2 shows a vector plot of the electric field in the $x-y$ plane located half way through the resonator and calculated at the Fano resonance at 10.8 μm. The circulating electric field seen in the inset is reminiscent of a magnetic dipole field pattern; however, rather than the usual in-plane magnetic dipole, the orientation of the resonant dipole is out of the plane of the array (i.e. a $z$–directed magnetic dipole). To further confirm this assignment, we performed the following "numerical experiment". First, we simulated the on-resonance response of the array (without a substrate for simplicity), placing a fictitious box around the center resonator of the array. Using Love's Equivalence Principle, we replace the sources within the box (i.e. the central resonator) with equivalent electric and magnetic surface currents derived from the total tangential fields on the box. We next remove all the other resonators in the array and calculate the fields radiated by the surface currents on the box. The radiated far-fields are then decomposed into their multipole components including all the quadrupole contributions [20]. Figure 3 shows the power radiated by the dominant four multipoles (the powers of the multipole components not shown in Fig. 3 are several orders of magnitude smaller than the dominant multipoles). Above and below the resonant frequency of ~27.8 THz (~10.8 μm), the $x$–directed electric dipole ($p_x$) dominates as expected. However, in the vicinity of the Fano resonance the strength of $p_x$ decreases dramatically. Simultaneously, the strength of the $z$–directed magnetic dipole ($m_z$) increases remarkably and dominates all other multipoles by nearly two orders of magnitude. This confirms the earlier assignment of the resonant mode as $m_z$. Note also that at the resonant frequency the $y$–directed electric dipole ($p_y$) and the $z$–$x$ magnetic quadrupole ($M_{zx}$) are also excited. However, the fields radiated by $p_y$ and $M_{zx}$ largely cancel each other in the forward and backward directions, which, in combination with the dramatic decrease of $p_x$, explains the high transmission and low reflection observed at the Fano resonance. The lack of perfect cancellation between $p_y$ and $M_{zx}$ results in a small depolarization of the transmitted wave (see the Supplementary material). The reflection spectrum shown in Fig. 2(b) exhibits an additional Fano resonance at shorter wavelengths. A similar analysis shows that the origin of the higher energy Fano resonance appearing at ~9.7 μm in Fig. 2(b) is analogous to the mechanism described above. However, in this case the bright mode is an in-plane magnetic dipole ($m_y$) and the dark mode is a longitudinal electric dipole ($p_z$). This second Fano resonance will not be discussed further in this paper.

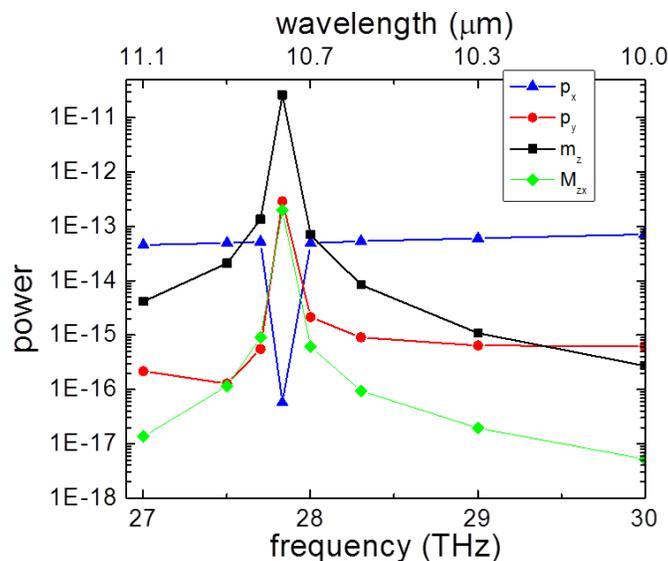

Fig. 3. The power (in W) radiated by the four dominant multipoles of the Fano resonators when excited in the array (no substrate). At the resonant frequency the $x$–directed electric dipole is largely extinguished and the $z$–directed magnetic dipole dominates.

We believe that the excitation of the $m_z$ multipole arises due to the different widths (in the $x$–direction) of the two parts of the resonator. Considering each part of the resonator as a separate polarizable dielectric region, we see that such an approximate spatial decomposition indicates the two separated electric dipoles will exhibit slightly different dipole strengths. Upon excitation with an $x$–polarized wave, the asymmetry of the two dipoles will lead to a $z$–directed magnetic field in the vicinity of the center of the resonator which can couple to and excite the $m_z$ dipole. Such an excitation mechanism is unavailable for the symmetric full cube resonator.

The large Q-factors of the Fano resonances arise due to the small radiative and non-radiative decay rates of the $z$-directed magnetic dipole in the array. For an isolated resonator, the $z$-directed dipole is free to radiate and is also subject to non-radiative decay processes arising from material absorption. This results in broad resonance linewidths for the isolated resonator. In contrast, when placed in the two-dimensional array, the resonator's normal radiative decay is compensated by driving terms arising from the local field at the position of the resonator [21, 22], leaving only the (small) non-radiative processes. Thus, the overall Q-factor of the resonators, and hence the Q-factor of the Fano resonance, becomes large. To demonstrate the importance of array effects in establishing the Fano resonance, we simulated the response of finite sized arrays (no substrate for simplicity) of varying sizes (see Fig. 4). For the isolated resonator, no Fano resonance is observed and the electric field vector plots are reminiscent of a px excitation. For the other simulations, the frequency of the Fano resonance shifts slightly as the array size increases. The 3 x 3 array shows a very weak Fano resonance, and the on-resonance electric field vector plots are complicated but begin to show field circulation within each resonator. The 5 x 5 array exhibits a clear Fano resonance, and the vector field plots for the interior resonators clearly show the electric field circulation associated with the mz dipole. Interestingly, the innermost resonator of the array shows the largest field enhancement, while the resonators at the edge of the array (which experience a drastically different local field and can radiate substantially more) show smaller field enhancements and less well defined modes. Proceeding to the 7 x 7 and 9 x 9 arrays, the number of interior resonators experiencing large field enhancements increases with array size, and once again the outermost resonators show weaker excitations. The absolute magnitude of the field enhancement for the centermost resonator rises sharply with array size for the smaller arrays, and is beginning to saturate at the largest (9 x 9) array (see Fig. 5). A magnetic field enhancement on the order of ~200 is observed. Thus, the overall size of the array required to achieve a robust Fano resonance is quite small.

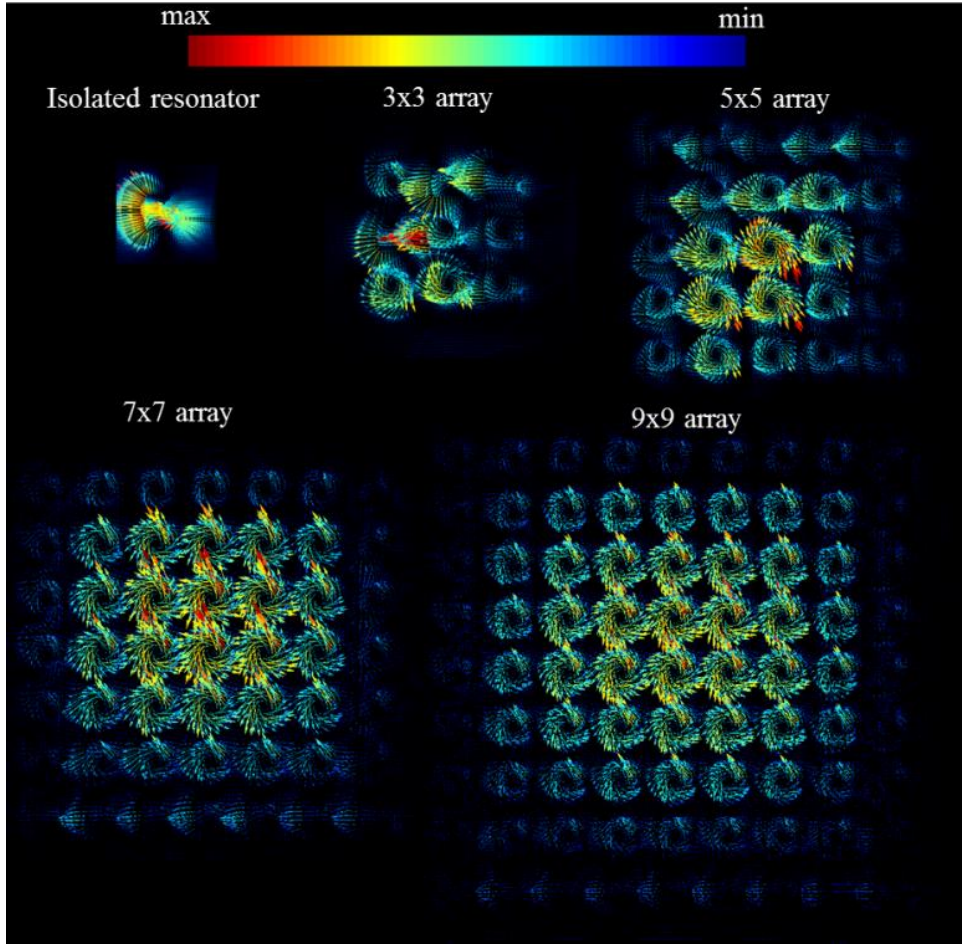

Fig. 4: Vector plots of the on-resonance electric field for the isolated resonator and 3 x 3, 5 x 5, 7 x 7, and 9 x 9 arrays. The vector plots were obtained at the vertical mid-plane of the resonators. The magnitude of the electric field is scaled so that each array image has the same scaling. The color bar maximum and minimum have been adjusted for each array since the absolute magnitude of the electric field increases sharply with the array size (see the Supplementary material).

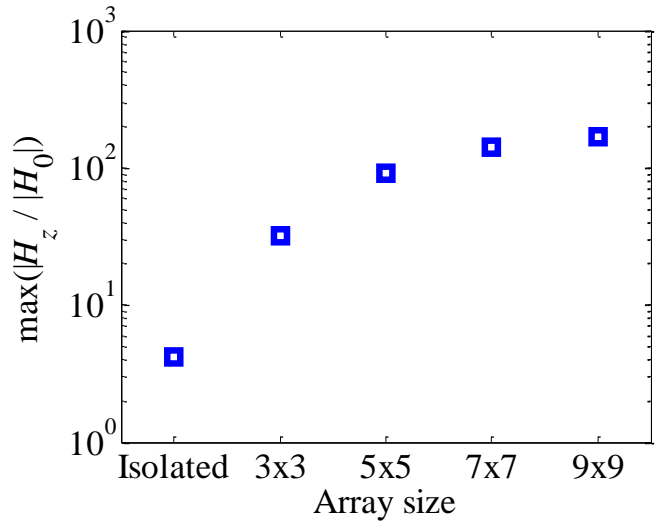

Fig. 5: The absolute magnitude of the magnetic field enhancement for the centermost resonator rises sharply with array size for the smaller arrays, and is beginning to saturate at the largest (9 x 9) array. The overall size of the array required to achieve a robust Fano resonance is quite small.

As a first experimental demonstration of the new metasurface design, we have designed and fabricated silicon-based Fano resonators operating near 1 μm wavelength. The resonators were fabricated using silicon-on-insulator wafers with a 250 nm thick silicon layer. The nominal side length of the resonators is 280 nm and the array spacing is 550 nm (see the Supplementary material for the design details). The arrays were fabricated using e-beam lithography, and reactive ion etching and reflectivity spectra were measured using a custom built near-infrared polarizing microscope coupled to a high-resolution spectrometer equipped with a CCD array detector. The inset of Fig. 6(a) shows a scanning electron micrograph (SEM) of several of the resonators in the array, along with the experimentally measured reflectivity spectrum. The measured spectrum is similar in features to the simulated spectrum (reported in the Supplementary material ), although an overall wavelength shift is observed. We attribute this shift to slight dimensional differences between the designed and fabricated arrays. The measured quality-factor of the resonance at 998.8 nm is about 350. This sharp feature is quite robust to the angle of incidence: we have measured with several objectives with NA of 0.14, 0.26, and 0.4 and observed the Fano resonances in all cases. Furthermore, there is a slight difference between the simulated and measured spectra in the shape of the shorter wavelength Fano resonance that we believe is due to a small inaccuracy in the SOI buried oxide thickness used in the simulations. The buried oxide forms a low finesse etalon that interferes with the Fano behavior and modulates the overall shape of the resonance.

For a second, even more compelling demonstration of the new broken symmetry Fano approach, we fabricated a GaAs-based Fano metasurface by adapting a processing scheme originally developed for surface emitting semiconductor lasers [23, 24]. In contrast to indirect bandgap Si used in the experimental demonstration described above, GaAs features a direct bandgap so that residual absorptive losses should be smaller in the near-infrared spectral range and larger Q-values might be possible. The resonator arrays were fabricated using epitaxially-grown GaAs layers and employ a novel means of isolating the resonators from the native GaAs substrate on which they were grown. Details of the resonator design are presented in the Supplementary material. Figure 6(b) shows the experimental reflectivity spectrum for three GaAs Fano metasurfaces with slightly different in-plane dimensional scaling factors, s, of a nominal 270 x 270 x 300 nm$^3$ (s = 1.0) design (see the Supplementary material for details). As expected, the Fano resonances shift to longer wavelength as the scaling factor increases. Notably, the FWHM of the Fano resonance is 1.6 nm for the metasurface corresponding to the largest scaling factor. This corresponds to a Q-factor of ~600, which is to our knowledge the largest metasurface Q-factor achieved in this spectral range for any type of metasurface presented in the literature. Low temperature reflectivity measurements (not shown here) showed a shifting of the spectral location of the Fano resonance, but did not reveal any further narrowing of the resonance. The achievement of such high-Q resonances along with their large field enhancements in GaAs is particularly exciting since it opens up new avenues for device designs that exploit the active and nonlinear properties of GaAs.

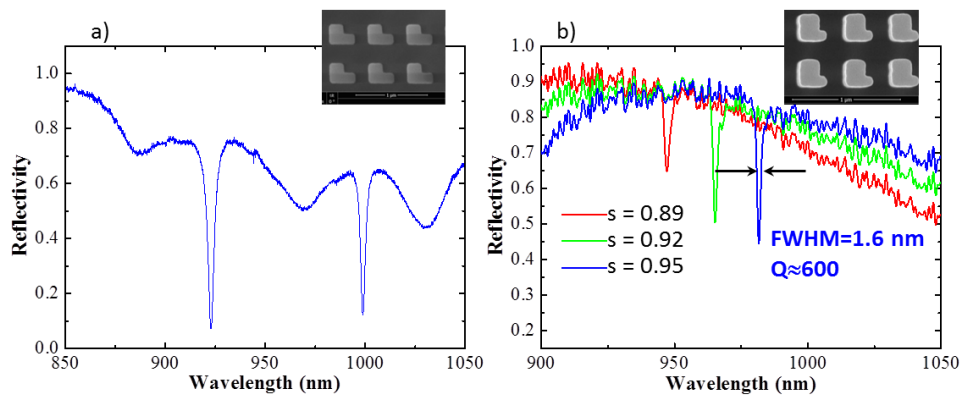

Fig. 6. (a) The experimentally measured reflectivity spectrum of the Si resonator array shown in the inset. The Q-factor for the resonance at 998.8 nm is ~350. (b) The measured reflectivity spectrum from three GaAs-based Fano metasurfaces with slightly different in-plane scalings (s) of a nominal 270 x 270 x 300 nm3 design. The array corresponding to the largest scaling factor exhibits a resonance with a FWHM of 1.6 nm, corresponding to a Q-factor of approximately 600. The inset shows an SEM of several of the resonators in the s = 0.95 array.

In conclusion, we have shown how breaking the symmetry of otherwise symmetric resonators leads to a new Fano metasurface design that features only one resonator in the unit cell and produces very high Q-factor resonances in the metasurface transmission and reflection spectra. The new approach is scalable over wide frequency ranges and can be easily implemented in semiconducting materials. In addition we have presented two experimental demonstrations of the new design operating in the near-infrared (~1 μm). The first utilizes silicon-on-insulator and achieves a Q-factor of ~350. The second demonstration uses GaAs resonators and achieves a Q-factor of ~600, the largest Q-factor ever reported using metasurfaces in this spectral range. We anticipate that the extension of our Fano metasurface approach to GaAs will allow us to exploit the active and nonlinear properties of GaAs to develop new devices such as lasers, photodetectors, modulators, nonlinear frequency converters, and sensors.

**Acknowledgment**. The authors acknowledge fruitful discussions with Prof. Edward Kuester, University of Colorado Boulder. Parts of this work were supported by the U.S. Department of Energy, Office of Basic Energy Sciences, Division of Materials Sciences and Engineering and performed, in part, at the Center for Integrated Nanotechnologies, an Office of Science User Facility operated for the U.S. Department of Energy (DOE) Office of Science. Portions of this work were supported by the Laboratory Directed Research and Development program at Sandia National Laboratories. Sandia National Laboratories is a multi-program laboratory managed and operated by Sandia Corporation, a wholly owned subsidiary of Lockheed Martin Corporation, for the U.S. Department of Energy's National Nuclear Security Administration under contract DE-AC04-94AL85000.

# Supplementary Material for "Broken symmetry dielectric resonators for high quality-factor Fano metasurfaces"

Salvatore Campione,[1,2,3,*] Sheng Liu,[1,2,3] Lorena I. Basilio,[1] Larry K. Warne,[1] William L. Langston,[1] Ting S. Luk,[1,2] Joel R. Wendt,[1] John L. Reno,[1,2] Gordon A. Keeler,[1] Igal Brener,[1,2] Michael B. Sinclair,[1,*]
[1]Sandia National Laboratories, P.O. Box 5800, Albuquerque, NM 87185, USA
[2]Center for Integrated Nanotechnologies, Sandia National Laboratories, P.O. Box 5800, Albuquerque, NM 87185, USA
[3]S.C. and S.L. contributed equally
*Corresponding authors: sncampi@sandia.gov or mbsincl@sandia.gov

This Supplementary material provides the dimensional details of the Ge, Si, and GaAs resonator designs; the small depolarization of the transmitted wave; and the simulated reflectivity of the experimental SOI design.

## 1. Dimensional details of the Ge, Si, and GaAs resonator designs.

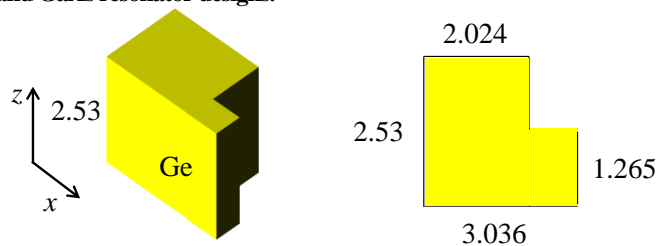

Fig. S1: Dimensional details in µm of the Ge resonators. The period of the metasurface is 4.2 µm.

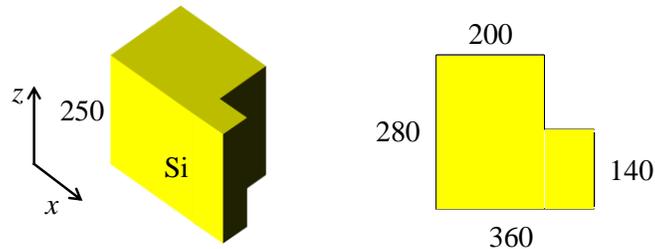

Fig. S2: Dimensional details in nm of the Si resonators. The period of the metasurface is 550 nm.

We designed three GaAs Fano resonators with the same array pitch of 470 nm and height of 300 nm but different in-plane unit cell dimensional scaling factors of 0.89, 0.92, and 0.95. Figure 2 shows a schematic of the unit cell with a scaling factor of 1 that comprises a cuboid with nominal side length of 270 nm and a smaller cuboid notch with side dimensions of $x = 70$ nm and $y = 190$ nm cutting through the resonator.

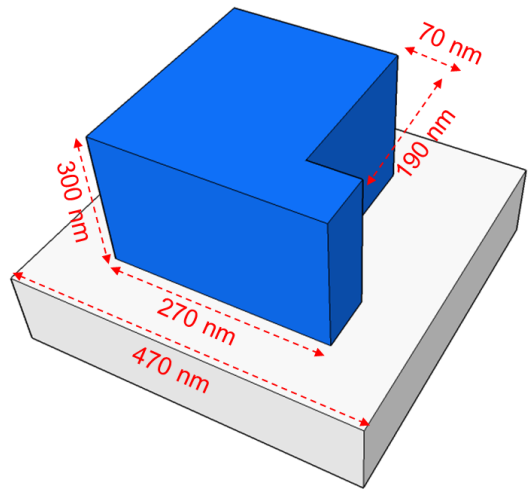

Fig. S3: Dimensional details in nm of the GaAs resonators. The period of the metasurface is 470 nm.

## 2. Small depolarization of the transmitted wave.

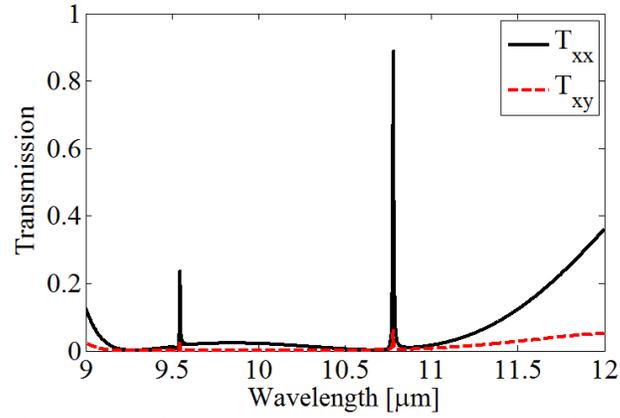

Fig. S4: The *x*- and *y*-polarized transmission spectrum of the Ge-based array in Fig. 2 obtained using full-wave simulations under *x*-polarized incidence. Because of the lack of perfect cancellation between $p_y$ and $M_{zx}$ mentioned in Fig. 3, a small depolarization of the transmitted wave is observed.

## 3. Simulated reflectivity of the experimental SOI design.

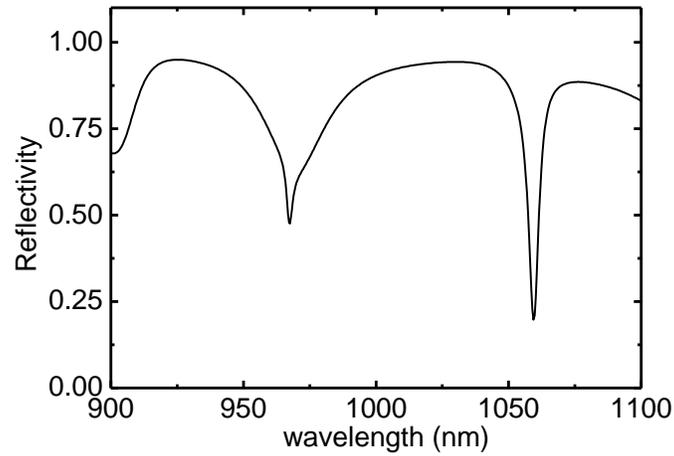

Fig. S5: The reflectivity spectrum of the SOI-based array in Fig. 6(a) obtained using full-wave simulations under *x*-polarized incidence. A good agreement with the experiment is observed.